\documentclass[aps,preprint]{revtex4-1}%
\usepackage{amssymb}
\usepackage{amsfonts}
\usepackage{amsmath}
\usepackage{graphicx}%
\begin{document}
\title{Frame Transformations for Fermions}
\author{Chris W Patterson}
\affiliation{Los Alamos National Laboratory, Los Alamos NM 87545}
\author{William G Harter}
\affiliation{Physics Department, University of Arkansas, Fayetteville AR 72701}

\begin{abstract}
\textit{The analog to the Legendre addition theorem is found for half-integral
angular momentum using frame transformations for rotor states.}

\end{abstract}
\maketitle

\section{Introduction}

\qquad The Legendre addition theorem for spherical harmonics applies to the
coupling of two particles with equal orbital angular momentum $\ell$ to give
total orbital angular momentum of zero. If we denote the spherical angles of
particle one by $(1)=(\theta_{1},\phi_{1})$, and of particle two by
$(2)=(\theta_{2},\phi_{2})$, then the addition theorem for spherical harmonics becomes%

\begin{equation}
\sqrt{4\pi/(2\ell+1)}\sum\limits_{m}(-1)^{m}\psi_{-m}^{l}(1)\psi_{m}%
^{l}(2)=\psi_{0}^{l}(\overline{1}),\label{1.1}%
\end{equation}
where $(\overline{1})\equiv(\overline{\theta}_{1},\overline{\phi}_{1})$ are
the body coordinates of particle one with respect to particle two. The
addition theorem represents a transformation from lab coordinates $(1)$ and
$(2)$ to body coordinates $(\overline{1})$ called a \textit{frame
transformation. }The addition theorem is symmetric with respect to exchange of
coordinates $(1)$ and $(2)$ or, equivalently, $(\overline{1})$ and
$(\overline{2})$. Because the total angular momentum in the lab frame is zero,
the left and right hand sides of the equation are rotational invariants.

\qquad It is interesting that the only attribution of this addition theorem to
Legendre found by the authors in the current literature is in the book by
Whittaker and Watson which was initially published in 1902 [1,2]. Today, the
addition theorem for spherical harmonics is derived without any reference
given to Legendre because it is so widely known and so easily proven. For
example, the well known book by Rose [3] on angular momentum theory gives two
proofs of the addition theorem without references.

\qquad When the two coupled orbital angular momenta are not equal, so that the
total angular momentum is not zero, a generalized addition theorem or frame
transformation can still be derived for rotor states as shown by Chang and
Fano [4]. Indeed, the addition theorem of Legendre is a special case of the
frame transformation for rotor states. Frame transformations are used to
transform from a weakly coupled basis in which two particles move nearly
independently to a strongly coupled one in which one of the particles
`follows' the other. In effect, we transform from the lab frame in the weakly
coupled case to the body frame in the strongly coupled case. Frame
transformations were used by Chang and Fano for diatomic molecules in which an
electron was weakly coupled to the molecular frame at large distances but
strongly coupled nearby. Their frame transformations theory was put on a firm
group theoretical footing using the representation theory of rotors with the
systematic treatment of inversions and molecular symmetries by Harter, et al
[5]. We will follow the latter treatment here. However, in this work what is
called the strongly coupled basis was called the Born-Oppenheimer Approximate
(BOA) basis in [5]. Frame transformation theory uses the standard matrix
relations of rotor states given by Casimir [6]. For a detailed exposition of
rotor representation theory in physics, we refer the reader to the books by
Biedenharn and Louck [7] and Harter[8] and the references therein {.}

\qquad Surprisingly, after nearly two hundred years, there is no analogous
addition theorem when the coupled angular momenta are equal and half-integral
which occurs for spin one-half particles. That is, in the above equation, what
happens when we replace integral $\ell$ by half-integral $j$ so that the
wavefunctions are no longer spherical harmonics? The left hand side of
(\ref{1.1}) will still be a rotational invariant. However, on the right hand
side $\psi_{0}^{j}(\overline{1})$ is not defined because $j$ is half-integral
and the body-axis projection is half-integral as well and cannot be zero. Is
there, then, an analogous addition theorem for half-integral$~j$?

\qquad The purpose of this work is to extend the addition theorem for
spherical harmonics and the frame transformations for rotor states to the case
of half-integral angular momentum coupling. This would correspond to the
coupling of two particles which have both integral orbital angular momentum
$\ell$ and intrinsic spin one-half angular momentum $%
\frac12
$. That is, we will consider the angular momentum coupling of two fermions in
the $j-j$ coupling limit which is useful for atomic and nuclear shell theory.
We give frame transformations for the general case of the coupling of unequal
half-integral angular momentum states and then the corresponding addition
theorem for the coupling of equal half-integral angular momentum states which
is special case of these frame transformations. Below, in order to make this
presentation self-contained, we first rederive the spin zero frame
transformation relations of Chang and Fano [4] and the addition theorem of
Legendre [1], using the formalism of Harter, et al [5] before proceeding to
their spin one-half analogs.

\section{Spin Zero Particles (Bosons)}

\subsection{Weakly Coupled Basis}

\qquad We let particle one have orbital angular momentum $\mathbf{l}_{1}$ and
particle two have orbital angular momentum $\mathbf{l}_{2}$ and assume that
the interaction between them is such that the total angular momentum
$\mathbf{L=l}_{1}\mathbf{+l}_{2}$ is conserved. We may write the total angular
momentum wavefunction as a product wavefunction of the spherical harmonics
using the Wigner coupling coefficients such that%

\begin{equation}
\Psi_{\ell_{1}\ell_{2}}^{LM}(weak)=\sum_{m_{1},m_{2}}C_{m_{1}m_{2}M}^{\ell
_{1}\text{ }\ell_{2}\text{ }L}\text{ }\psi_{m_{1}}^{\ell_{1}}(1)\psi_{m_{2}%
}^{\ell_{2}}(2), \label{2.1}%
\end{equation}
where the condition $m_{1}+m_{2}=M$ on such sums is understood implicitly. We
call this wavefunction the weakly coupled basis which is appropriate if the
two particles are not strongly coupled. However, if the two particles are
strongly coupled so that one particle `follows' the other, then it is better
to transform to the moving frame of one of the particles, say, particle two.
Such a transformation is called a frame transformation. In order to effect
such a transformation it is necessary to use rotor wavefunctions.

\subsection{Strongly Coupled Basis and Rotor States}

\qquad We let particle two define the body frame so that the body \={z}-axis
is at spherical angles $(\theta_{2},\phi_{2})$ with respect to the lab z-axis
and corresponds to Euler angles $(\theta_{2},\phi_{2},\gamma_{2})$ where
$\gamma_{2}$ is an arbitrary rotation about the body \={z}-axis. We use the
standard rotor wavefunctions [5-8] such that%

\begin{equation}
\psi_{m_{2}}^{\ell_{2}}(2)\equiv\sqrt{(2\ell_{2}+1)/4\pi}D_{m_{2}0}^{l_{2}%
\ast}(2). \label{2.2}%
\end{equation}
We may now transform the particle one wavefunction into this body frame using%

\begin{equation}
\psi_{m_{1}}^{\ell_{1}}(1)=\sum_{m_{1}^{\prime}}D_{m_{1}m_{1}^{\prime}}%
^{l_{1}\ast}(2)\psi_{m_{1}^{\prime}}^{\ell_{1}}(\overline{1}).\label{2.3}%
\end{equation}
Note that the Euler angle $\gamma_{2}$ is superfluous in (\ref{2.2}) but not
in (\ref{2.3}). Physically in (\ref{2.3}) $m_{1}$ is the projection of the
angular momentum along the lab z-axis, whereas $m_{1}^{\prime}$ is the
projection of the angular momentum along the body \={z}-axis as defined by
particle two. In (\ref{2.2}) we have $m_{2}^{\prime}=0$, so the projection of
the angular momentum along the body \={z}-axis is zero. This is necessary
because a point particle can have no orbital angular momentum about an axis
through it. The same would be true if (\ref{2.2}) represented the wavefunction
about the axis of a diatomic molecule as was the case considered by Chang and
Fano [4].

\qquad We may now rewrite (\ref{2.1}) so that%

\begin{equation}
\Psi_{\ell_{1}\ell_{2}}^{LM}(weak)=\sqrt{(2\ell_{2}+1)/4\pi}\sum_{m_{1}%
,m_{2},m_{1}^{\prime}}C_{m_{1}m_{2}M}^{\ell_{1}\text{ }\ell_{2}\text{ }%
L}\text{ }D_{m_{1}m_{1}^{\prime}}^{l_{1}\ast}(2)D_{m_{2}0}^{l_{2}\ast}%
(2)\psi_{m_{1}^{\prime}}^{\ell_{1}}(\overline{1}).\label{2.4}%
\end{equation}
Using the relations,%

\begin{equation}
\sum_{m_{1},m_{2}}C_{m_{1}m_{2}M}^{\ell_{1}\text{ }\ell_{2}\text{ }L}\text{
}D_{m_{1}m_{1}^{\prime}}^{l_{1}\ast}(2)D_{m_{2}0}^{l_{2}\ast}(2)=C_{m_{1}%
^{\prime}0\text{ }m_{1}^{\prime}}^{\ell_{1}\text{ }\ell_{2}\text{ }L}%
D_{Mm_{1}^{\prime}}^{L\ast}(2), \label{2.5}%
\end{equation}
we find the \textbf{frame transformations for spin zero particles}:%

\begin{align}
\Psi_{\ell_{1}\ell_{2}}^{LM}(weak) &  =\sqrt{(2\ell_{2}+1)/4\pi}\sum
_{m_{1}^{\prime}}C_{m_{1}^{\prime}0\text{ }m_{1}^{\prime}}^{\ell_{1}\text{
}\ell_{2}\text{ }L}D_{Mm_{1}^{\prime}}^{L\ast}(2)\psi_{m_{1}^{\prime}}%
^{\ell_{1}}(\overline{1})\label{2.6}\\
&  =\sqrt{(2L+1)/4\pi}\sum_{m_{1}^{\prime}}(-1)^{l_{1}-m_{1}^{\prime}%
}C_{m_{^{1}}^{\prime}-m_{1}^{\prime}0}^{L\text{ }\ell_{1}\text{ }\ell_{2}%
}D_{Mm_{1}^{\prime}}^{L\ast}(2)\psi_{m_{1}^{\prime}}^{\ell_{1}}(\overline
{1})\nonumber\\
&  =\sum_{m_{1}^{\prime}}(-1)^{l_{1}-m_{1}^{\prime}}C_{m_{^{1}}^{\prime}%
-m_{1}^{\prime}0}^{L\text{ }\ell_{1}\text{ }\ell_{2}}\Psi_{\ell_{1}\ell
_{2}m_{1}^{\prime}}^{LM}(strong),\nonumber
\end{align}
where we define the strongly coupled basis to be%

\begin{equation}
\Psi_{\ell_{1}\ell_{2}m_{1}^{\prime}}^{LM}(strong)=\sqrt{(2L+1)/4\pi}%
D_{Mm_{1}^{\prime}}^{L\ast}(2)\psi_{m_{1}^{\prime}}^{\ell_{1}}(\overline
{1}).\label{2.7}%
\end{equation}
In the strongly coupled basis we couple the total angular momentum
$\mathbf{L}$ to that of particle one $\mathbf{l}_{1}$ to get the orbital
angular momentum of particle two $\mathbf{l}_{2}$. This is equivalent to
angular momentum subtraction where $\mathbf{L-l}_{1}=\mathbf{l}_{2}$. The
rotor wavefunction $D_{Mm_{1}^{\prime}}^{L\ast}(2)$ has total angular momentum
quantum number $L$ and carries the particle one orbital angular momentum
projection $m_{1}^{\prime}$ along its body \={z}-axis.

\qquad If we let the total angular momentum be zero, $L=0$, so that
$M=m_{1}^{\prime}=0$ and $\ell_{1}=\ell_{2}\equiv\ell$, we find%

\[
\sum_{m}C_{-m\text{ }m\text{ }0}^{\ell\text{ }\ell\text{ }0}\text{ }\psi
_{-m}^{l}(1)\psi_{m}^{l}(2)=(-1)^{\ell}\psi_{0}^{\ell}(\overline{1}%
)/\sqrt{4\pi},
\]
or the addition theorem for spherical harmonics, namely, the \textbf{Legendre
addition theorem for spin zero particles}:%

\begin{equation}
\sqrt{4\pi/(2\ell+1)}\sum\limits_{m}(-1)^{m}\psi_{-m}^{l}(1)\psi_{m}%
^{l}(2)=\psi_{0}^{l}(\overline{1}).\label{2.8}%
\end{equation}
Note that while in the lab frame the total angular momentum $L$ is zero, in
the body frame the total angular momentum is $l$. The total angular momentum
is not conserved in the body frame because it is not an inertial frame.
However, the projection of the angular momentum along both the lab z-axis and
the body \={z}-axis is zero. We now wish to derive the equivalent addition
theorem for a system in which both particles have spin one-half in addition to
their orbital angular momenta. Below, we will derive the frame transformations
for total angular momentum $J$ and then let $J=0$ to derive the analogous
addition theorem for two spin one-half particles.

\section{Spin One-Half Particles (Fermions)}

\subsection{Weakly Coupled Basis}

\qquad We now let the two particles have spin one-half in addition to their
angular momentum using the $j-j$ coupling scheme. Particle one has total
angular momentum $\mathbf{j}_{1}\mathbf{=l}_{1}\mathbf{+%
\frac12
}$, particle two has total angular momentum $\mathbf{j}_{2}\mathbf{=l}%
_{2}\mathbf{+%
\frac12
}$, and the total angular momentum of the two particles is $\mathbf{J=j}%
_{1}\mathbf{+j}_{2}$. The weakly coupled basis using $j-j$ coupling now becomes%

\begin{align}
\Psi_{j_{1}j_{2}}^{JN\ell_{1}\ell_{2}}(weak)  &  =\sum_{n_{1,}n_{2}}%
C_{n_{1}\text{ }n_{2}\text{ }N}^{j_{1}\text{ }j_{2}\text{ }J}\psi_{n_{1}%
}^{j_{1}}(1)\psi_{n_{2}}^{j_{2}}(2)\label{3.1}\\
&  =\sum_{n_{1},n_{2}}\text{ \ }\sum_{m_{1},m_{2},\sigma_{1},\sigma_{2}%
}C_{n_{1}\text{ }n_{2}\text{ }N}^{j_{1}\text{ }j_{2}\text{ }J}C_{m_{1}\text{
}\sigma_{1}\text{ }n_{1}}^{\ell_{1}\text{ }%
\frac12
\text{ \ }j_{1}}C_{m_{2}\text{ }\sigma_{2}\text{ }n_{2}}^{\ell_{2}\text{ }%
\frac12
\text{ \ }j_{2}}\psi_{m_{1}}^{l_{1}}(1)\chi_{\sigma_{1}}^{%
\frac12
}(1)\psi_{m_{2}}^{l_{2}}(2)\chi_{\sigma_{2}}^{%
\frac12
}(2).\nonumber
\end{align}
We could equally well have used $L-S$ coupling instead of $j-j$ coupling to
prove the relations below, although the derivation is more difficult. Note
that in the sums above the sums on $m_{1},\sigma_{1}$ are restricted by the
condition $m_{1}+\sigma_{1}=n_{1}$ and the sums on $m_{2},\sigma_{2}$ are
restricted by the condition $m_{2}+\sigma_{2}=n_{2}$.

\subsection{Strongly Coupled Basis and Rotor States}

\qquad We now transform to the body \={z}-axis of particle two. In addition to
(\ref{2.2}) and (\ref{2.3}) for transformations of the orbital wavefunctions,
we also have the following transformations for the spin wavefunctions:%

\begin{equation}
\chi_{\sigma_{2}}^{%
\frac12
}(2)=\sum_{\sigma_{2}^{\prime}}D_{\sigma_{2}\sigma_{2}^{\prime}}^{%
\frac12
\ast}(2)\chi_{\sigma_{2}^{\prime}}^{%
\frac12
}(\overline{2}), \label{3.2}%
\end{equation}
and%

\begin{equation}
\chi_{\sigma_{1}}^{%
\frac12
}(1)=\sum_{\sigma_{1}^{\prime}}D_{\sigma_{1}\sigma_{1}^{\prime}}^{%
\frac12
\ast}(2)\chi_{\sigma_{1}^{\prime}}^{%
\frac12
}(\overline{1}). \label{3.3}%
\end{equation}
where $(\overline{2})=(0,0)$ are the body coordinates of particle two with
respect to the particle two axis. The spin functions $\chi_{\sigma_{2}}^{%
\frac12
}(2)$ in (\ref{3.2}) have the proper normalization as shown below. It is
interesting to compare (\ref{2.2}) with (\ref{3.2}). In (\ref{2.2}) we see
that the body component of $\psi_{m_{2}}^{\ell_{2}}(2)$ is $m_{2}^{\prime}=0$,
whereas in (\ref{3.2}) we see that the body component of $\chi_{\sigma_{2}}^{%
\frac12
}(2)$ is a linear combination of the possible components $\sigma_{2}^{\prime
}=-%
\frac12
$ and $\sigma_{2}^{\prime}=+%
\frac12
$. This means that, physically, there is an equal likelihood of finding
$\sigma_{2}^{\prime}$ to be $-%
\frac12
$ or $+%
\frac12
$ about an arbitrary body axis. Indeed, we may use (\ref{2.2}) and (\ref{3.2})
as the actual definitions of $\psi_{m_{2}}^{\ell_{2}}(2)$ and $\chi
_{\sigma_{2}}^{%
\frac12
}(2)$, respectively, in order to properly specify both the lab and body
components of orbital and spin angular momentum. In this sense any other
definition would be incomplete.

\qquad Using the relations%

\begin{equation}
\sum_{m_{1},\sigma_{1}}C_{m_{1}\text{ }\sigma_{1}\text{ }n_{1}}^{\ell
_{1}\text{ }%
\frac12
\text{ \ }j_{1}}D_{m_{1}m_{1}^{\prime}}^{\ell_{1}\ast}(2)D_{\sigma_{1}%
\sigma_{1}^{\prime}}^{%
\frac12
\ast}(2)=C_{m_{1}^{\prime}\text{ }\sigma_{1}^{\prime}\text{ }n_{1}^{\prime}%
}^{\ell_{1}\text{ }%
\frac12
\text{ \ }j_{1}}D_{n_{1}n_{1}^{\prime}}^{j_{1}\ast}(2), \label{3.4}%
\end{equation}
where $m_{1}^{\prime}+\sigma_{1}^{\prime}=n_{1}^{\prime}$ and%

\begin{equation}
\sum_{m_{2},\sigma_{2}}C_{m_{2}\text{ }\sigma_{2}\text{ }n_{2}}^{\ell
_{2}\text{ }%
\frac12
\text{ \ }j_{2}}D_{m_{2}0}^{\ell_{2}\ast}(2)D_{\sigma_{2}\sigma_{2}^{\prime}%
}^{%
\frac12
\ast}(2)=C_{0\text{ }\sigma_{2}^{\prime}\text{ }\sigma_{2}^{\prime}}^{\ell
_{2}\text{ }%
\frac12
\text{ \ }j_{2}}D_{n_{2}\sigma_{2}^{\prime}}^{j_{2}\ast}(2), \label{3.5}%
\end{equation}
we find%

\begin{align*}
\Psi_{j_{1}j_{2}}^{JN\ell_{1}\ell_{2}}(weak) &  =\sqrt{(2\ell_{2}+1)/4\pi}%
\sum_{n_{1},n_{2,}\sigma_{2}^{\prime},n_{1}^{\prime}}C_{0\text{ }\sigma
_{2}^{\prime}\text{ }\sigma_{2}^{\prime}}^{\ell_{2}\text{ }%
\frac12
\text{ \ }j_{2}}[C_{n_{1}\text{ }n_{2}\text{ }N}^{j_{1}\text{ }j_{2}\text{ }%
J}D_{n_{1}n_{1}^{\prime}}^{j_{1}\ast}(2)D_{n_{2}\sigma_{2}^{\prime}}%
^{j_{2}\ast}(2)]\chi_{\sigma_{2}^{\prime}}^{%
\frac12
}(\overline{2})\\
&  \times\sum_{m_{1}^{\prime},\sigma_{1}^{\prime}}C_{m_{1}^{\prime}\text{
}\sigma_{1}^{\prime}\text{ }n_{1}^{\prime}}^{\ell_{1}\text{ }%
\frac12
\text{ \ }j_{1}}\psi_{m_{1}^{\prime}}^{\ell_{1}}(\overline{1})\chi_{\sigma
_{1}^{\prime}}^{%
\frac12
}(\overline{1}).
\end{align*}
It is important to note that the sum on $m_{1}^{\prime},\sigma_{1}^{\prime}$
is restricted by the condition $m_{1}^{\prime}+\sigma_{1}^{\prime}%
=n_{1}^{\prime}$. Using%

\[
C_{n_{1}^{\prime}\text{ }\sigma_{2}^{\prime}\text{ }(n_{1}^{\prime}+\sigma
_{2}^{\prime})}^{j_{1}\text{ \ }j_{2}\text{ \ \ }J}D_{N,n_{1}^{\prime}%
+\sigma_{2}^{\prime}}^{J\ast}(2)=\sum_{n_{1},n_{2}}C_{n_{1}\text{ }n_{2}\text{
}N}^{j_{1}\text{ }j_{2}\text{ }J}D_{n_{1}n_{1}^{\prime}}^{j_{1}\ast
}(2)D_{n_{2}\sigma_{2}^{\prime}}^{j_{2}\ast}(2),
\]
and%
\begin{equation}
\psi_{n_{1}^{\prime}}^{j_{1}}(\overline{1})\equiv\sum_{m_{1}^{\prime}%
,\sigma_{1}^{\prime}}C_{m_{1}^{\prime}\text{ }\sigma_{1}^{\prime}\text{ }%
n_{1}^{\prime}}^{\ell_{1}\text{ }%
\frac12
\text{ \ }j_{1}}\psi_{m_{1}^{\prime}}^{\ell_{1}}(\overline{1})\chi_{\sigma
_{1}^{\prime}}^{%
\frac12
}(\overline{1}),\label{3.5a}%
\end{equation}
we find the\textbf{\ frame transformation for spin one-half particles:}%

\begin{align}
\Psi_{j_{1}j_{2}}^{JN\ell_{1}\ell_{2}}(weak)  &  =\sqrt{(2\ell_{2}+1)/4\pi
}\sum_{\sigma_{2}^{\prime}}C_{0\text{ }\sigma_{2}^{\prime}\text{ }\sigma
_{2}^{\prime}}^{\ell_{2}\text{ }%
\frac12
\text{ \ }j_{2}}\sum_{n_{1}^{\prime}}C_{n_{1}^{\prime}\text{ }\sigma
_{2}^{\prime}\text{ }(n_{1}^{\prime}+\sigma_{2}^{\prime})}^{j_{1}\text{
\ }j_{2}\text{ \ \ }J}D_{N,n_{1}^{\prime}+\sigma_{2}^{\prime}}^{J\ast}%
(2)\psi_{n_{1}^{\prime}}^{j_{1}}(\overline{1})\chi_{\sigma_{2}^{\prime}}^{%
\frac12
}(\overline{2})\label{3.6}\\
&  =\sum_{\sigma_{2}^{\prime}}(-1)^{\sigma_{2}^{\prime}+%
\frac12
}C_{-\sigma_{2}^{\prime}\text{ }\sigma_{2}^{\prime}\text{ }0}^{j_{2}\text{ }%
\frac12
\text{ \ }\ell_{2}}\sum_{n_{1}^{\prime}}(-1)^{j_{1}-n_{1}^{\prime}}%
C_{(n_{1}^{\prime}+\sigma_{2}^{\prime})~-n_{1}^{\prime}~\sigma_{2}^{\prime}%
}^{J\text{ \ \ }j_{1}\text{ \ \ }j_{2}}\Psi_{j_{1}j_{2}n_{1}^{\prime}%
\sigma_{2}^{\prime}}^{JN\ell_{1}\ell_{2}}(strong),\nonumber
\end{align}
where we define the strongly coupled basis to be%

\begin{equation}
\Psi_{j_{1}j_{2}n_{1}^{\prime}\sigma_{2}^{\prime}}^{JN\ell_{1}\ell_{2}%
}(strong)=\sqrt{(2J+1)/4\pi}D_{N,n_{1}^{\prime}+\sigma_{2}^{\prime}}^{J\ast
}(2)\psi_{n_{1}^{\prime}}^{j_{1}}(\overline{1})\chi_{\sigma_{2}^{\prime}}^{%
\frac12
}(\overline{2}). \label{3.7}%
\end{equation}
The above should be compared with (\ref{2.6}) and (\ref{2.7}). In the strongly
coupled basis we couple the total angular momentum $\mathbf{J}$ to that of
particle one $\mathbf{j}_{1}$ to get the orbital angular momentum of particle
two $\mathbf{j}_{2}$ corresponding to angular momentum subtraction
$\mathbf{J-j}_{1}\mathbf{=j}_{2}$. We then couple the angular momentum
$\mathbf{j}_{2}$ to the spin of particle two to get the orbital angular
momentum of particle two $\mathbf{l}_{2}$ corresponding to angular momentum
subtraction $\mathbf{j}_{2}\mathbf{-%
\frac12
=l}_{2}$. The rotor state with total angular momentum quantum number $J$
carries the sum of angular momentum projections of particle one $n_{1}%
^{\prime}$ and particle two $\sigma_{2}^{\prime}$ along its body \={z}-axis,
where $n_{1}^{\prime}=m_{1}^{\prime}+\sigma_{1}^{\prime}$.

\qquad To derive the addition theorem for spin one-half particles we let $J=0
$, so that $N=0$ and $n_{1}^{\prime}=-\sigma_{2}^{\prime}$. Then from
(\ref{3.1}), (\ref{3.6}) and (\ref{3.7}) letting $j_{1}=j_{2}=j$ and
$n_{1}=-n_{2}=-n$, we find%

\[
\sum_{n}C_{-n~n~0}^{j~~j~~0}\psi_{-n}^{j}(1)\psi_{n}^{j}(2)=\sum_{\sigma
_{2}^{\prime}}(-1)^{j-%
\frac12
}C_{-\sigma_{2}^{\prime}\text{ }\sigma_{2}^{\prime}\text{ }0}^{j\text{ }%
\frac12
\text{ \ }\ell_{2}}C_{0~\sigma_{2}^{\prime}~\sigma_{2}^{\prime}}^{0~~j~~j}%
\psi_{-\sigma_{2}^{\prime}}^{j}(\overline{1})\chi_{\sigma_{2}^{\prime}}^{%
\frac12
}(\overline{2})/\sqrt{4\pi},
\]
or the \textbf{addition theorem for spin one-half particles}:%

\begin{equation}
\sqrt{4\pi/(2j+1)}\sum_{n}(-1)^{j+n}\psi_{-n}^{j}(1)\psi_{n}^{j}(2)=(-1)^{j-%
\frac12
}\left\{
\begin{array}
[c]{c}%
\lbrack\psi_{%
\frac12
}^{j}(\overline{1})\chi_{-%
\frac12
}^{%
\frac12
}(\overline{2})+\psi_{-%
\frac12
}^{j}(\overline{1})\chi_{%
\frac12
}^{%
\frac12
}(\overline{2})]/\sqrt{2}\text{ for }\ell_{2}=j+%
\frac12
\\
\lbrack\psi_{%
\frac12
}^{j}(\overline{1})\chi_{-%
\frac12
}^{%
\frac12
}(\overline{2})-\psi_{-%
\frac12
}^{j}(\overline{1})\chi_{%
\frac12
}^{%
\frac12
}(\overline{2})]/\sqrt{2}\text{ for }\ell_{2}=j-%
\frac12
\end{array}
\right\}  , \label{3.8}%
\end{equation}
where we have evaluated the Wigner coupling coefficients. This should be
compared to (\ref{2.8}) for spin zero particles. Keep in mind that $\psi_{%
\frac12
}^{j}(\overline{1})$ and $\psi_{-%
\frac12
}^{j}(\overline{1})$ on the right of (\ref{3.8}) refer to $j=j_{1}$ and are
given by (\ref{3.5a}). We may write these explicitly for $\ell_{1}=j+%
\frac12
,$%

\begin{align*}
\psi_{%
\frac12
}^{j}(\overline{1}) &  =\sqrt{\frac{2j+3}{4j+4}}\psi_{1}^{j+%
\frac12
}(\overline{1})\chi_{-%
\frac12
}^{%
\frac12
}(\overline{1})-\sqrt{\frac{2j+1}{4j+4}}\psi_{0}^{j+%
\frac12
}(\overline{1})\chi_{%
\frac12
}^{%
\frac12
}(\overline{1}),\\
\psi_{-%
\frac12
}^{j}(\overline{1}) &  =-\sqrt{\frac{2j+3}{4j+4}}\psi_{-1}^{j+%
\frac12
}(\overline{1})\chi_{%
\frac12
}^{%
\frac12
}(\overline{1})+\sqrt{\frac{2j+1}{4j+4}}\psi_{0}^{j+%
\frac12
}(\overline{1})\chi_{-%
\frac12
}^{%
\frac12
}(\overline{1}),
\end{align*}
and for $\ell_{1}=j-%
\frac12
,$%

\begin{align*}
\psi_{%
\frac12
}^{j}(\overline{1}) &  =\sqrt{\frac{2j-1}{4j}}\psi_{1}^{j-%
\frac12
}(\overline{1})\chi_{-%
\frac12
}^{%
\frac12
}(\overline{1})+\sqrt{\frac{2j+1}{4j}}\psi_{0}^{j-%
\frac12
}(\overline{1})\chi_{%
\frac12
}^{%
\frac12
}(\overline{1}),\\
\psi_{-%
\frac12
}^{j}(\overline{1}) &  =\sqrt{\frac{2j-1}{4j}}\psi_{-1}^{j-%
\frac12
}(\overline{1})\chi_{%
\frac12
}^{%
\frac12
}(\overline{1})+\sqrt{\frac{2j+1}{4j}}\psi_{0}^{j-%
\frac12
}(\overline{1})\chi_{-%
\frac12
}^{%
\frac12
}(\overline{1}).
\end{align*}
Thus (\ref{3.8}) actually consists of four different cases corresonding to
$\ell_{1}=j\pm%
\frac12
$ and $\ell_{2}=j\pm%
\frac12
.$ For interaction operators, such as the Coulomb interaction, which depend
only on the relative coordinates $(\overline{1})$ between the two particles,
it is much easier to evaluate matrix elements in $j-j$ coupling using the
strongly coupled basis on the right of (\ref{3.8}).

\qquad We can derive a special case of (\ref{3.8}) for pure spin states if we
let $\ell_{1}=\ell_{2}=0$ so that $j=%
\frac12
$, $\psi_{%
\frac12
}^{j}=\chi_{%
\frac12
}^{%
\frac12
}/\sqrt{4\pi}$, and $\psi_{-%
\frac12
}^{j}=\chi_{-%
\frac12
}^{%
\frac12
}/\sqrt{4\pi}$ on the left and right of (\ref{3.8}). The addition theorem for
spin states alone becomes:%
\begin{equation}
\lbrack\chi_{%
\frac12
}^{%
\frac12
}(1)\chi_{-%
\frac12
}^{%
\frac12
}(2)-\chi_{-%
\frac12
}^{%
\frac12
}(1)\chi_{%
\frac12
}^{%
\frac12
}(2)]/\sqrt{2}=[\chi_{%
\frac12
}^{%
\frac12
}(\overline{1})\chi_{-%
\frac12
}^{%
\frac12
}(\overline{2})-\chi_{-%
\frac12
}^{%
\frac12
}(\overline{1})\chi_{%
\frac12
}^{%
\frac12
}(\overline{2})]/\sqrt{2}. \label{3.9}%
\end{equation}
This same equation may also be derived by substituting (\ref{3.2}) and
(\ref{3.3}) directly on the left of (\ref{3.9}) and using the spin one-half
Wigner coupling coefficients. Note that from the definition of the spin
functions in (\ref{3.2}) we have
\begin{equation}
\psi_{\sigma}^{%
\frac12
}(2)=\chi_{\sigma}^{%
\frac12
}(2)/\sqrt{4\pi}=\sum_{\sigma^{\prime}}D_{\sigma\sigma^{\prime}}^{%
\frac12
\ast}(2)\chi_{\sigma^{\prime}}^{%
\frac12
}(\overline{2})/\sqrt{4\pi}, \label{3.10}%
\end{equation}
which has proper normalization for the rotational matrixes. The addition
theorem for identical fermions (\ref{3.8}) is antisymmetric with respect to
exchange of particles $(1)$ and $(2)$ or, equivalently, $(\overline{1})$ and
$(\overline{2})$.

References

[1] Legendre A M 1817 \textit{Calcul Int\'{e}gral}, \textbf{II} (Paris)

[2] Whittaker E T and Watson G N 1952 \textit{A Course of Modern Analysis}
(Cambridge University Press, Cambridge)

[3] Rose M E 1957 \textit{Elementary Theory of Angular Momentum} (John Wiley
and Sons, New York)

[4] Chang E S and Fano U 1972 Phys. Rev A \textbf{6} 173-185

[5] Harter W G, Patterson C W and da Paixao F J 1978 Rev. Mod. Phys.
\textbf{50} 37-83

[6] Casimir H B G 1931 \textit{Rotation of a Rigid Body in Quantum Mechanics}
(BIJ J.B. Wolters' Uitgevers-Maatschappij, Gronigen Batavia)

[7] Biedenharn L C and Louck J D 1981 Encyclopedia of Mathematics\textbf{\ 8},
\textit{Angular Momentum in Quantum Physic}s (Addison-Wesley, Reading)

[8] Harter W G 1993 \textit{Principles of Symmetry, Dynamics, and
Spectroscopy} (Wiley-Interscience, New York)
\end{document}